\documentstyle{article}
\begin{document}
\begin{titlepage}
\begin{flushright}
FERMILAB-Pub-96/332-T\\
\end{flushright}
\vspace{1cm}
\begin{center}
{\Large\bf Determination of $W$-boson Properties at Hadron Colliders}\\
\vspace{1cm}
{\large
W.~T.~Giele\\
\vspace{0.5cm}
and \\
\vspace{0.5cm}
S.~Keller}\\
\vspace{0.5cm}
{\it
Fermi National Accelerator Laboratory, P.~O.~Box 500,\\
Batavia, IL 60510, U.S.A.} \\
\vspace{0.5cm}
{\large April 1997}
\vspace{0.5cm}
\end{center}
\begin{abstract}
Methods for measuring the $W$-boson properties 
at hadron colliders are discussed.
It is demonstrated that the ratio 
between the $W$- and $Z$-boson 
observables can be reliably calculated 
using perturbative QCD, even when the
individual $W$- and $Z$-boson observables are not. 
Hence, by using a measured $Z$-boson observable 
and the perturbative calculation of the ratio 
of the $W$- over $Z$-boson observable, 
we can accurately predict the $W$-boson observable.
The use of the ratio reduces both the experimental and theoretical 
systematic uncertainties substantially.
Compared to the currently used methods
it might, at high luminosity, result 
in a smaller overall uncertainty on the 
measured $W$-boson mass and width.
\end{abstract}
\end{titlepage}

\section{Introduction}

The measurement of the $W$-boson mass and width at Hadron Colliders 
has a long history of slowly improving uncertainties. 
Now, with the expected high luminosity TEVATRON runs,
there will be a dramatic increase in the accuracy of 
the $W$-boson mass and width measurements. This will require a better 
understanding of the theoretical uncertainties 
resulting from uncalculated QED 
\cite{BeKl,BaKeWa} and QCD corrections. The latter 
is the subject of this paper. There are several problems with 
the theoretical predictions/models currently used in the extraction of
the $W$-boson properties. 

First of all, the phenomenological models used by the experimenters 
combine certain measured aspects of vector boson production with 
the theoretical calculation. 
The D0 collaboration~\cite{D0W} uses the resummed calculation of
Ref.~\cite{LaYu}. 
The theoretical model used is based on 
Next-to-Leading Order (NLO) matrix elements \cite{AR}
in which the large logarithms associated with the
small values of the vector boson transverse momentum,
$P_T^V$, are resummed \cite{AK}.
This resummation technique \cite{CoSoSt}
necessitates the input of experimental
data to determine the non-perturbative phenomenological parameters.
In Ref.~\cite{LaYu}, lower energy Drell-Yan 
data were used for this purpose.
The D0 collaboration uses its own Drell-Yan data
to determine the non-perturbative parameters.
Once the non-perturbative
parameters have been determined the resummed calculation is used 
to predict the leptonic observables from which the
$W$-boson properties are extracted.
The CDF collaboration \cite{CDFW} uses a leading-order 
calculation of the $W$-boson production and decay
which is subsequently folded with the $P_T^W$-spectrum.    
The $Z$-boson transverse momentum distribution is used as 
a first guess for the $P_T^W$-spectrum and is subsequently 
scaled~\cite{CDFWFULL} in order 
to match the component of the recoil energy perpendicular to the 
direction of the charged lepton in $W$-boson events. 
In both the D0 and CDF procedure it is difficult 
to evaluate the theoretical uncertainty.

Secondly, certain observables like the transverse 
energy distribution of the charged lepton are not
perturbatively calculable due to large 
radiative corrections. This forces the experimenters to 
rely more on the transverse mass distribution 
to extract the $W$-boson mass.
This distribution does not suffer from large radiative corrections 
(i.e., is not very sensitive to the $P_T^W$-spectrum), 
but by definition
depends on the  neutrino transverse 
momentum.  The reconstruction of the latter  
requires the measurement of the hadronic part of the 
event, resulting in substantial systematic uncertainties 
which ultimately 
will limit the precision on the $W$-boson mass and width 
determination. 
This problem of missing energy reconstruction will 
be compounded in the future TEVATRON runs because 
of the expected increase in the number of interactions 
per crossing~\cite{TeV33}.

The obvious solution to both of the above problems is not new,
but with the planned high luminosity TEVATRON runs it
can be brought to full maturity. 
The basic idea is to use the measured 
lepton distributions in $Z$-boson decay,
along with the calculated ratio of the $W$- over $Z$-boson
distribution,
to predict the equivalent leptonic distribution in the $W$-boson case. 
By comparing this prediction to the measured leptonic distribution the 
$W$-boson mass and width can be extracted.
Because of the use of the ratio, the $W$-boson mass and width will
be given relative to the $Z$-boson mass and width.
The $Z$-boson mass and width are accurately 
known from the LEP experiments \cite{lepZ}.
This method reduces many of the experimental systematic 
uncertainties because they are largely correlated
between the $W$- and $Z$-boson distributions.
The drawback is that the 
leptonic $Z$-boson cross section is an order
of magnitude smaller than the leptonic $W$-boson cross section, 
resulting in a larger statistical uncertainty. However, 
the exchange of the systematic uncertainties for 
statistical uncertainties is
exactly what is needed for future high luminosity runs. 
This method also reduces the theoretical uncertainty as
only the ratio of the distribution has to be calculated, 
and not the distribution themselves.  As we will see, 
because $W$- and $Z$-boson productions 
properties are very similar, the large radiative 
corrections that might affect the 
individual distributions cancel in the ratio.  
In other word, the ratio can be reliably calculated 
using perturbative QCD even in regions of phase space
where the individual distributions cannot.  
For example, all the large logarithms associated 
with small values of the transverse 
energy of the vector boson cancel in the ratio and we
get a well behaved perturbative expansion which has a 
solid footing in the theory (see e.g. Ref.~\cite{Reno} 
for a comparison between the transverse momentum of the
$W$- and $Z$-boson).  There is no need for resummation in the 
calculation of the ratio.    
The theoretical uncertainty associated with perturbative QCD 
can be estimated from the size of the next-to-leading order 
calculation compared to the leading order calculation.
If necessary, the uncertainty could be further decreased by 
calculating the next-to-next-to-leading order 
corrections \cite{Neerven}. 
This gives a good understanding of the theoretical 
uncertainty on the $W$-boson mass and width measurements,
which will be important in order to interpret the upcoming
high luminosity results at the TEVATRON.

One remaining unknown at the moment is the uncertainty
related to the parton distribution functions. Obviously, by taking
the $W$- over $Z$-boson ratio these uncertainties will also be reduced.  
The ratio will
strongly depend on the $u$-quark/$d$-quark ratio. 
However, with the current status of 
Parton Density Functions (PDF's) there is no way of 
quantifying this statement~\cite{SnoPDF}.

In section 2, we set up the theoretical framework 
for the comparison between the $W$- and $Z$-boson observables. 
In section 3 we consider explicitly the transverse momentum of 
the vector boson.  The transverse mass and the transverse energy 
of the charged lepton distributions are discussed in sections 
4 and 5, respectively.
Finally, in section 6 we summarize our findings and 
draw some conclusions.

\section{The theoretical framework}

The basic idea is to use the $Z$-boson observables 
combined with perturbative QCD to predict 
the $W$-boson observables.
The main difference between the $W$- and $Z$-boson
production is due to the 
difference in the mass.  To minimize this difference we 
consider the ratio of the observable ${\cal O}$, 
which has the dimension of mass, and the vector boson
mass $M_V$:
\begin{equation}
X_{{\cal O}}^V = \frac{{\cal O}^V}{M_V}\ .
\end{equation}
Next, we define the ratio
\begin{equation}
R_{{\cal O}}(X_{{\cal O}}) = \frac{A_W(X_{{\cal O}}^W=X_{{\cal O}})}
                                  {A_Z(X_{{\cal O}}^Z=X_{{\cal O}})}\ ,
\end{equation}
where $A_V$ is the differential cross section with respect 
to the scaled variable:
\begin{equation}
A_V\left(X_{{\cal O}}^V\right)\ =
\frac{d\,\sigma_V}{d\,X_{\cal O}^V}. 
\end{equation}
The measured $Z$-boson differential cross section, 
$A_Z(X_{\cal O})|_{\mbox{measured}}$, can be used along with the calculated
ratio $R_{\cal O}(X_{\cal O})$ 
to predict the $W$-boson differential cross section,
$A_W(X_{\cal O})|_{\mbox{predicted}}$:
\begin{equation}
A_W(X_{{\cal O}})|_{\mbox{predicted}} = R_{{\cal O}}(X_{{\cal O}})\times
A_Z(X_{{\cal O}})|_{\mbox{measured}}.
\end{equation}
This expression can be used to relate the differential cross section
with respect to the non-scaled variables:
\begin{equation}
\left.\frac{d\,\sigma^W}{d\,{\cal O}^W}\right|_{\mbox{predicted}} =
\left.\frac{M_Z}{M_W}\times
R_{{\cal O}}\left( \frac{{\cal O}^W}{M_W} \right)\times
\frac{d\,\sigma^Z}{d\,{\cal O}^Z}\right|_{\mbox{measured}}
^{{\cal O}^Z=\frac{M_Z}{M_W}{\cal O}^W}\ .
\label{eq:WvsZ}
\end{equation}
Finally we define the ``$K$-factor'' as the ratio 
of the Next-to-Leading Order (NLO)
over Leading Order (LO) calculation
of $R_{\cal O}( X_{\cal O} )$:
\begin{equation}
K_{\cal O}(X_{\cal O}) = \frac{R_{\cal O}^{NLO}(X_{\cal O})}{R_{\cal O}^{LO}(X_{\cal O})}.
\end{equation}
To demonstrate that the ratio can be calculated
perturbatively, we will show that for the observables of interest
the $K$-factor is close to unity over the relevant $X_{\cal O}$ range.
In other words we will show that the scaled
$W$- and $Z$-boson distributions have 
very similar radiative corrections.  

In the explicit calculation of the ratio we use the 
DYRAD\footnote{The program can be obtained from the WWW-page:
``WWW-theory.fnal.gov/people/giele/dyrad.html''.}
program~\cite{dyrad}.  For the renormalization and factorization 
scales we choose the average of the $W$- and $Z$-boson masses. 
Other choices can be made for the scales, but the dependence 
of the ratio on that choice is small.
Similar cuts should be imposed on the $Z$- and $W$-boson decay leptons.
The transverse energy lepton cuts should be 
applied on the mass scaled variables 
to avoid large radiative corrections in the region close to the cuts.  
The only cut that cannot 
be matched between the $W$- and $Z$-boson case 
is the rapidity cut on the second charged
lepton in the $Z$-boson decay.
This is because the longitudinal momentum of the neutrino 
in the $W$-boson decay cannot be reconstructed. 
However, this difference is geometrical in nature and 
any difference will be modeled accurately by the perturbative calculation.
On the other hand, the rapidity cut on the second charged lepton 
in the $Z$-boson decay should be as weak as possible 
in order to increase the $Z$-boson sample.  
In this paper we will not impose cuts on the 
leptons as they have no baring on the 
argumentation or the conclusions 
reached.  It is, however, trivial to 
include these cuts using the DYRAD program.   
The leptonic branching ratio's were 
not included in any of the numerical results.

\section{The transverse momentum of the vector boson}

The transverse momentum of the vector boson, $P_T^V$,
is an interesting test 
of the ratio method for several reasons.
First of all, for both the $W$- and $Z$-boson there are published 
measurements of the transverse vector boson momentum. This means we can
actually apply the method described in section 2 and predict the $W$-boson
$P_T$-distribution using the $Z$-boson $P_T$-distribution
 over the entire $P_T$-spectrum with a small
theoretical uncertainty. The current data sets of both CDF and D0 
contain an integrated luminosity of well over 
100 pb$^{-1}$. Unfortunately, the vector boson transverse momentum results
using this high statistics data have not been published yet.
We will use the CDF $W$-boson data containing 
an integrated luminosity of 4.1 pb$^{-1}$ \cite{CDFWPt}
and the D0 $Z$-boson data using an integrated luminosity 
of 12.8 pb$^{-1}$ \cite{D0ZPt} to demonstrate the
potential of the ratio method.
Secondly, the transverse $P_T$-distribution has an
infrared unstable region as the
transverse momentum goes to zero and it is interesting 
to study the behavior of the ratio in that region.

The general form of the $n$-th order scaled transverse momentum 
differential cross section is \cite{AK}
\begin{equation}
A_V^{(n)}(X_{P_T}) = \frac{\sigma_V^{(0)}}{X_{P_T}}
\sum_{k=1}^{n}\sum_{m=0}^{2k-1} a_{k,m}^V \alpha_S^k \log^m(X_{P_T})
+R(\alpha_S)\ ,
\end{equation} 
where $X_{P_T}=P_T^V/M_V$, $R(\alpha_S)$ represent the remaining terms
for which the divergence is weaker than $1/X_{P_T}$,
$\sigma_V^{(0)}$ is the born cross section and $\alpha_S$
is the strong coupling constant.  
In the limit that $X_{P_T}\rightarrow 0$ all terms 
in the expansion can be neglected with respect to the leading 
logarithmic term: 
\begin{equation}
\lim_{X_{P_T}\rightarrow 0} A^{(n)}_V(X_{P_T}) = \frac{a_{n,2n-1}}{X_{P_T}}
\alpha_S^n\log^{2n-1}(X_{P_T}) \sigma_V^{(0)},
\end{equation}
where the constants $a_{n,2n-1}$ cannot depend on the vector boson type
because of the universality of the leading logarithms, such that
\begin{equation}
\lim_{X_{P_T}\rightarrow 0} R^{(n)}_{P_T} = 
\frac{\sigma_W^{(0)}} {\sigma_Z^{(0)}}.
\label{eq:limit1}
\end{equation}
At any order in $\alpha_S$ the ratio converges toward the leading order result 
in the infrared unstable region, 
i.e. the radiative corrections to the ratio are very small in that region.
There is no need for resummation of large logarithms,
the perturbation series is well behaved
and the theoretical uncertainty is small.
This despite the fact that the individual differential 
cross section in the numerator and denominator 
of the ratio are unreliable at small $X_T$.

\begin{figure}[t]\vspace{10cm}
\includegraphics{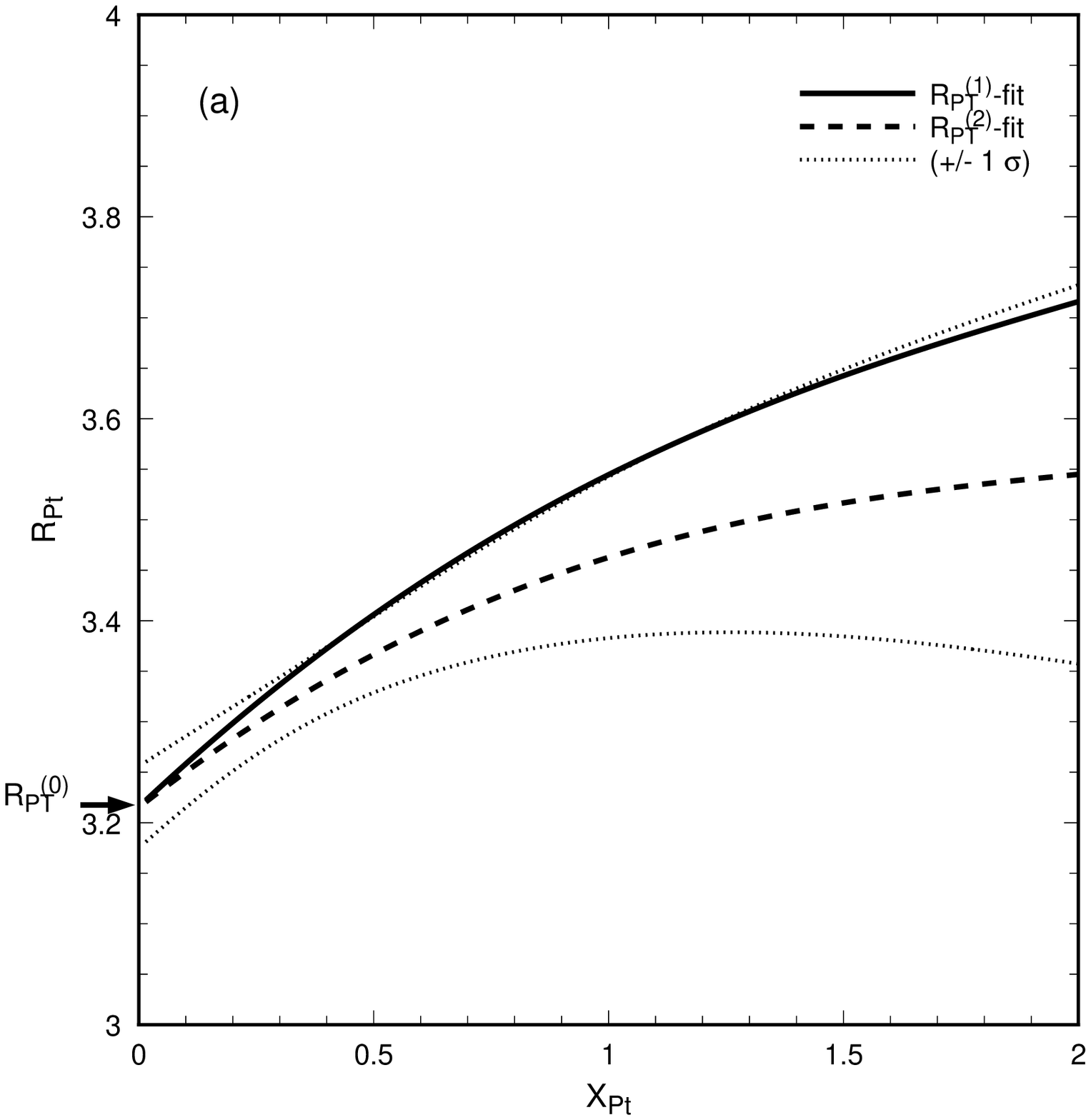}
\includegraphics{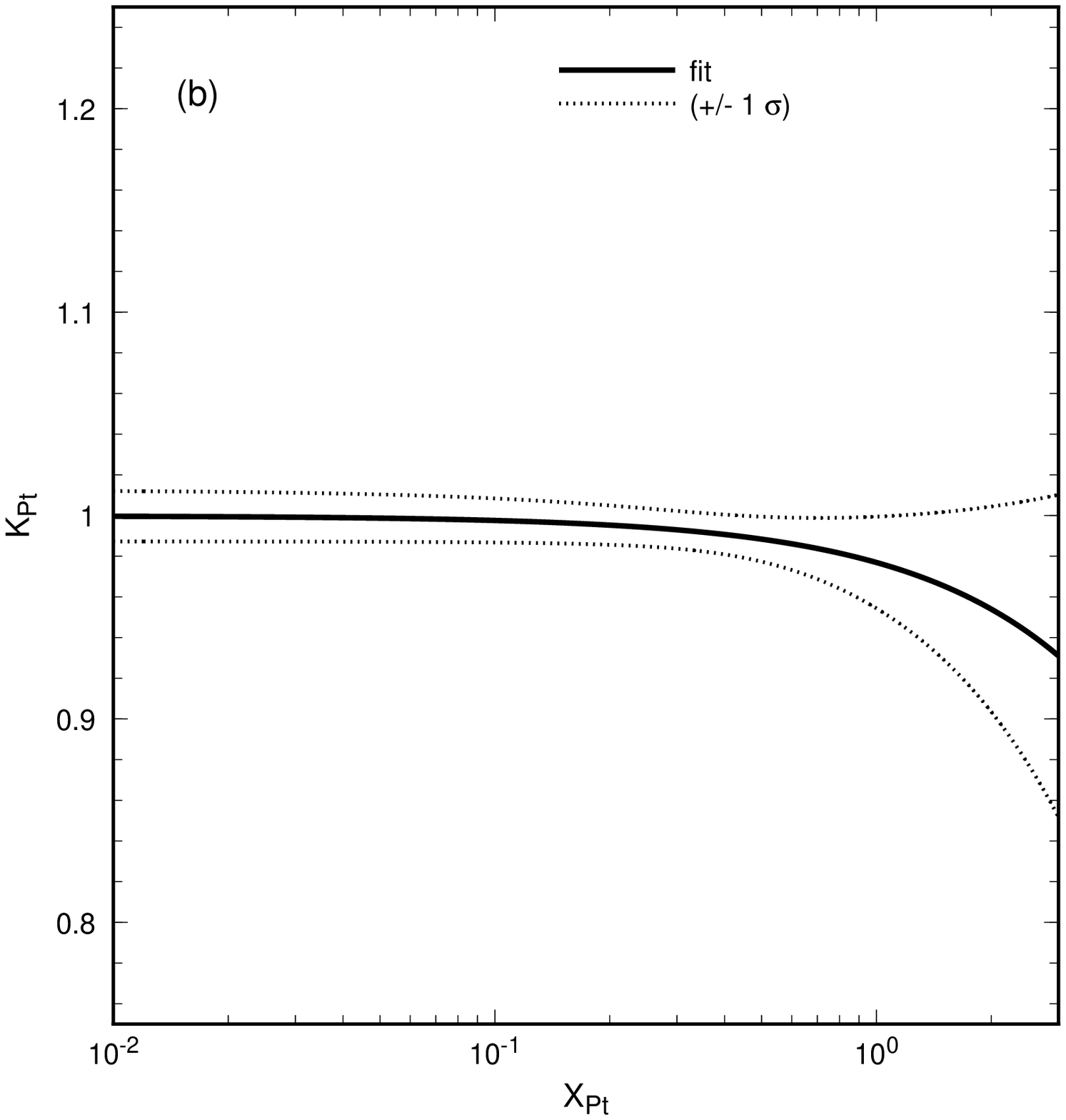}
\caption[]
{a) The ratio $R_{P_T}$ as a function of the scaled vector
 boson transverse momentum, $X_{P_T}=P_T^V/M_V$, at different orders in
$\alpha_S$.    The arrow on the left side represent
the LO value of the ratio.  The solid line and the dashed lines are the result
of fits to the order $\alpha_S$ and $\alpha_S^2$ Monte-Carlo calculations.   
The dotted lines represent the one sigma uncertainty
due to the Monte-Carlo integration for the 
order $\alpha_S^2$ calculation.  The uncertainty
for the order $\alpha_S$ is smaller and not shown.
b) The $K$-factor, $K_{P_T}=R_{P_T}^{(2)}/R_{P_T}^{(1)}$, 
as a function of $X_{P_T}$ together with
the uncertainty range associated with the Monte-Carlo integration.}
\end{figure}
\begin{figure}[t]\vspace{10cm}
\includegraphics{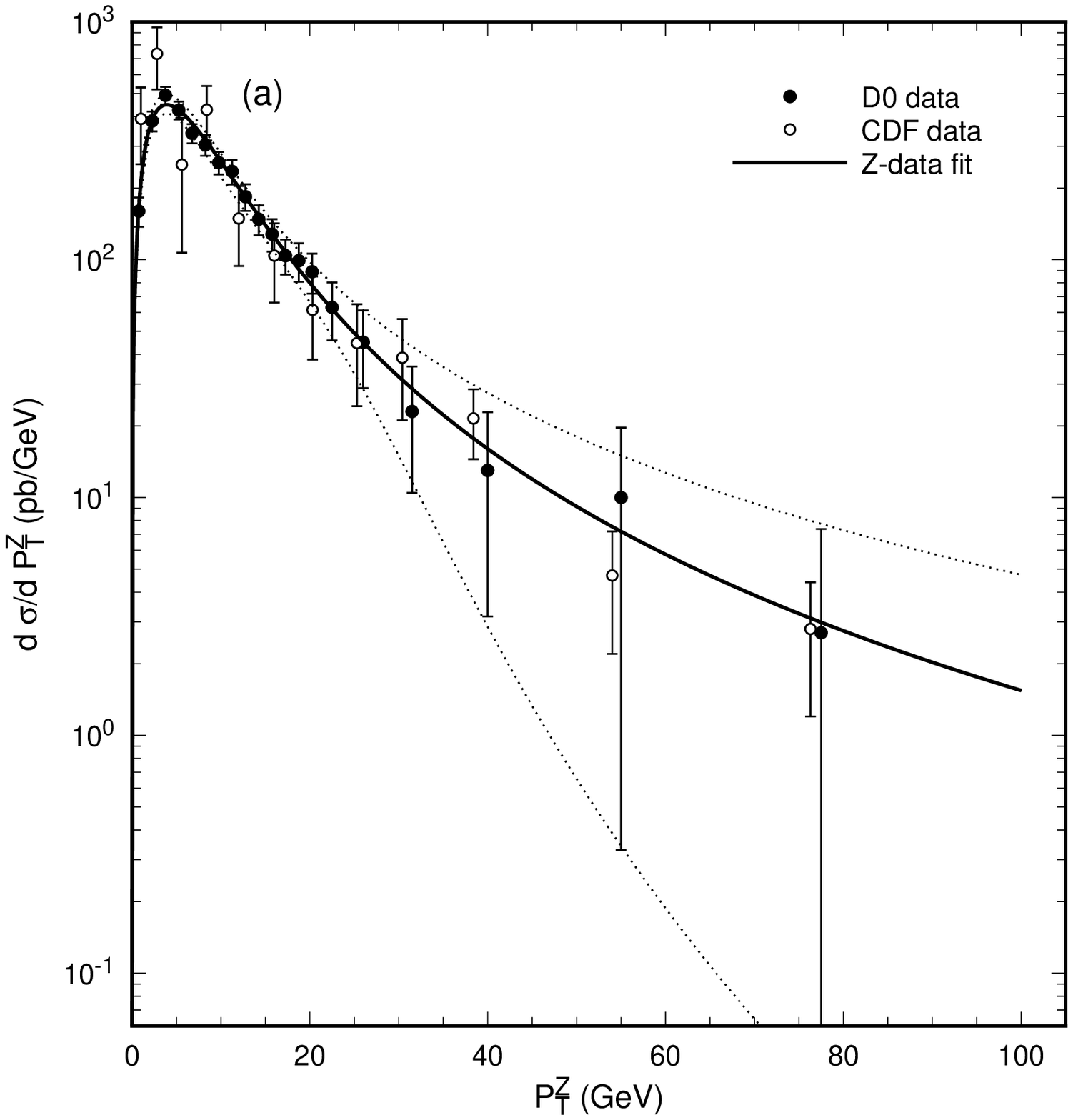}
\includegraphics{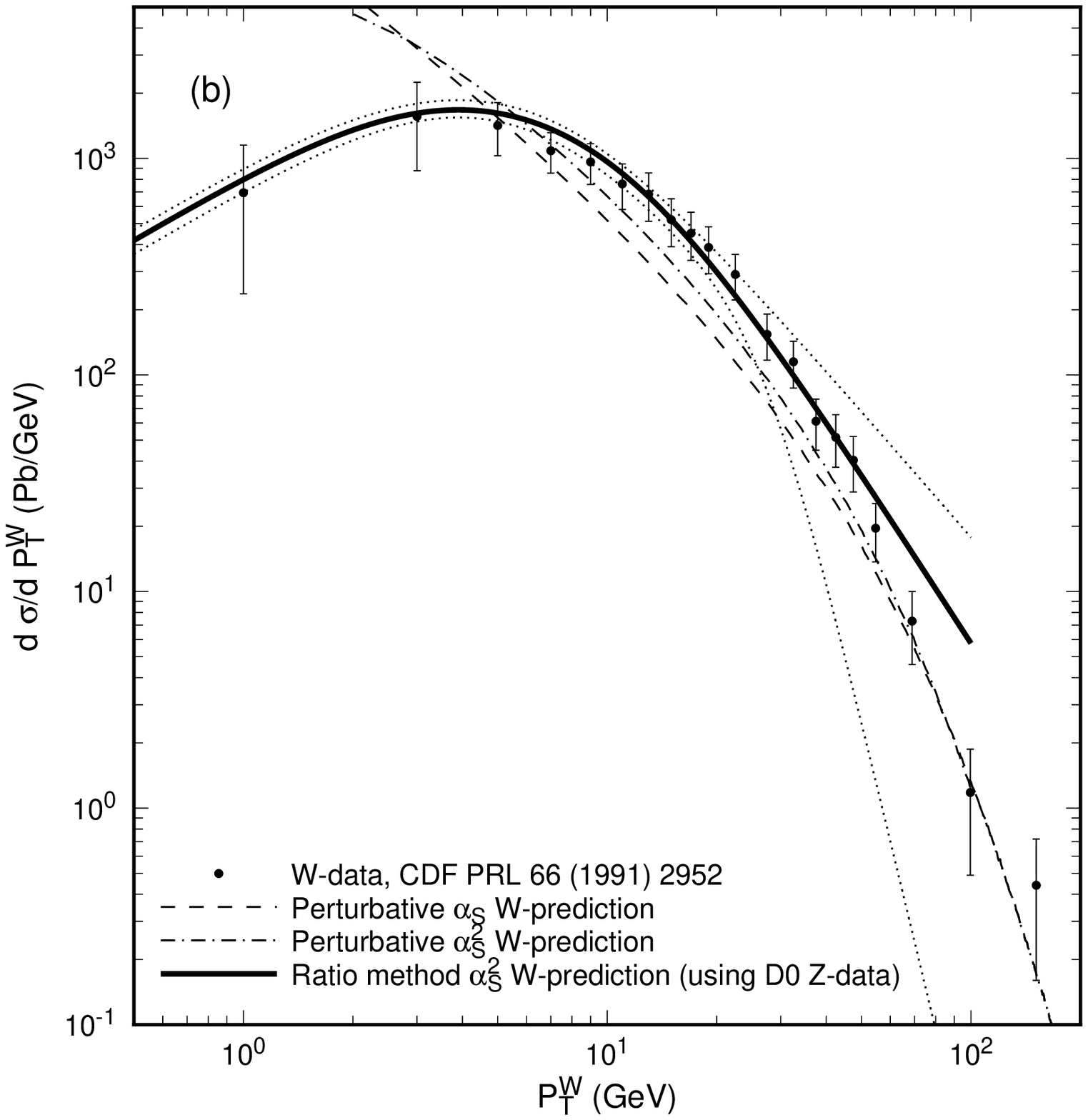}
\caption[]
{a) The $P_T^Z$-distributions for both the 
CDF~\cite{CDFZPt} and D0~\cite{D0ZPt} data.
 The solid line represent the fit to the D0-data as given in ref. \cite{D0ZPt}.
The dotted lines represent a fit to the D0-data uncertainties. 
b) The $W$-boson CDF-data, along with the perturbative 
order $\alpha_S$ and $\alpha_S^2$ calculation and the order 
$\alpha_S^2$ prediction using the ratio method with the D0 $Z$-boson data.
The dotted lines represent the uncertainties stemming from the D0-data.}
\end{figure}
\begin{figure}[t]\vspace{10cm}
\includegraphics{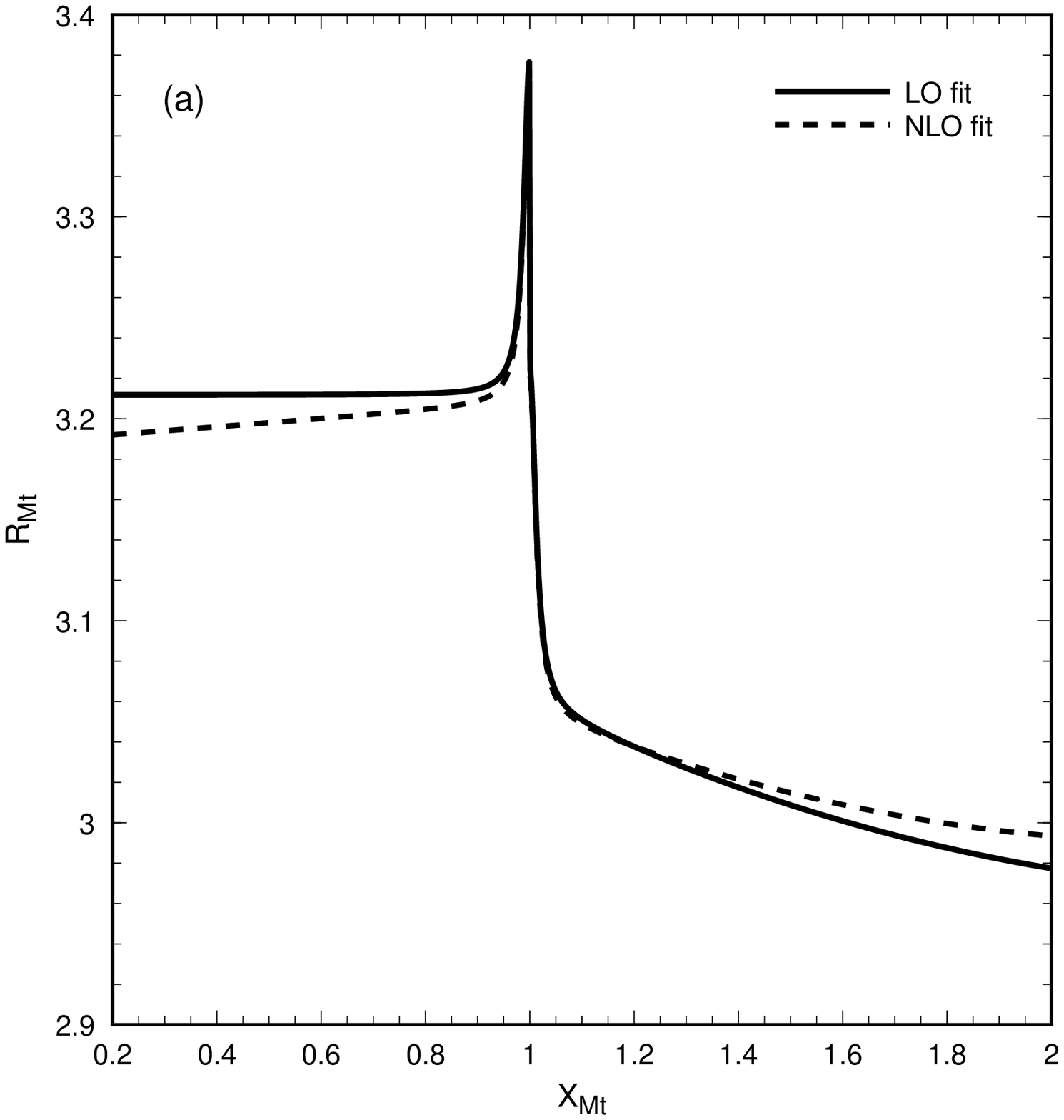}
\includegraphics{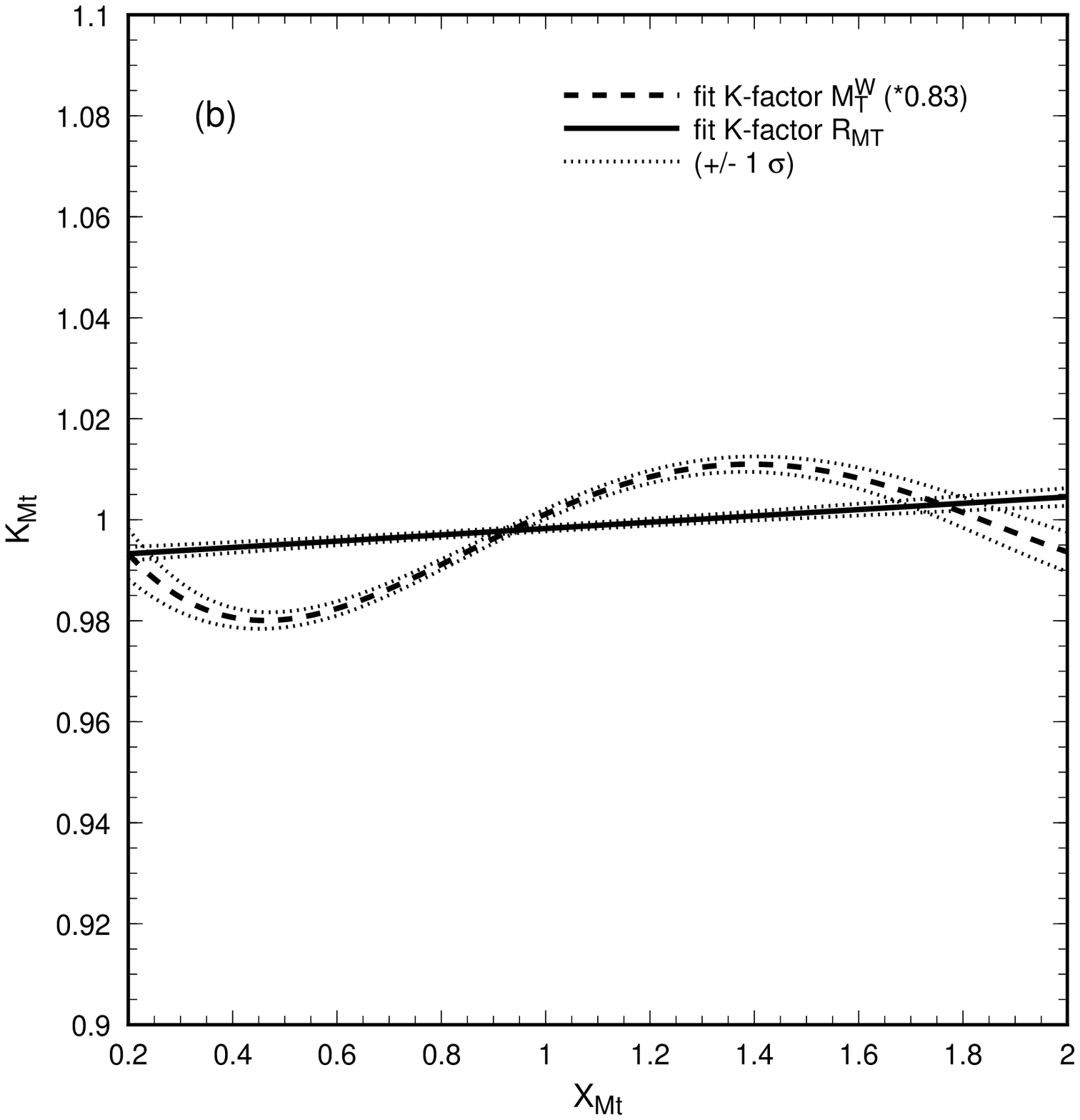}
\caption[]
{a) The LO (solid line) and NLO (dashed line) ratio $R_{M_T}$ as a 
function of the scaled transverse mass $X_{M_T}=M_T^V/M_V$.   
The leptonic branching fractions are not included.
b) The $K$-factor $K_{M_T}=R_{M_T}^{(1)}/R_{M_T}^{(0)}$ 
as a function of $X_{M_T}$ 
(solid line), and the K-factor for the $W$ transverse mass (dashed line), 
normalized to 1 at $X_{MT}=1$.  In both case we have included 
the one sigma uncertainty range associated 
with the Monte-Carlo integration.}
\end{figure}
\begin{figure}[t]\vspace{10cm}
\includegraphics{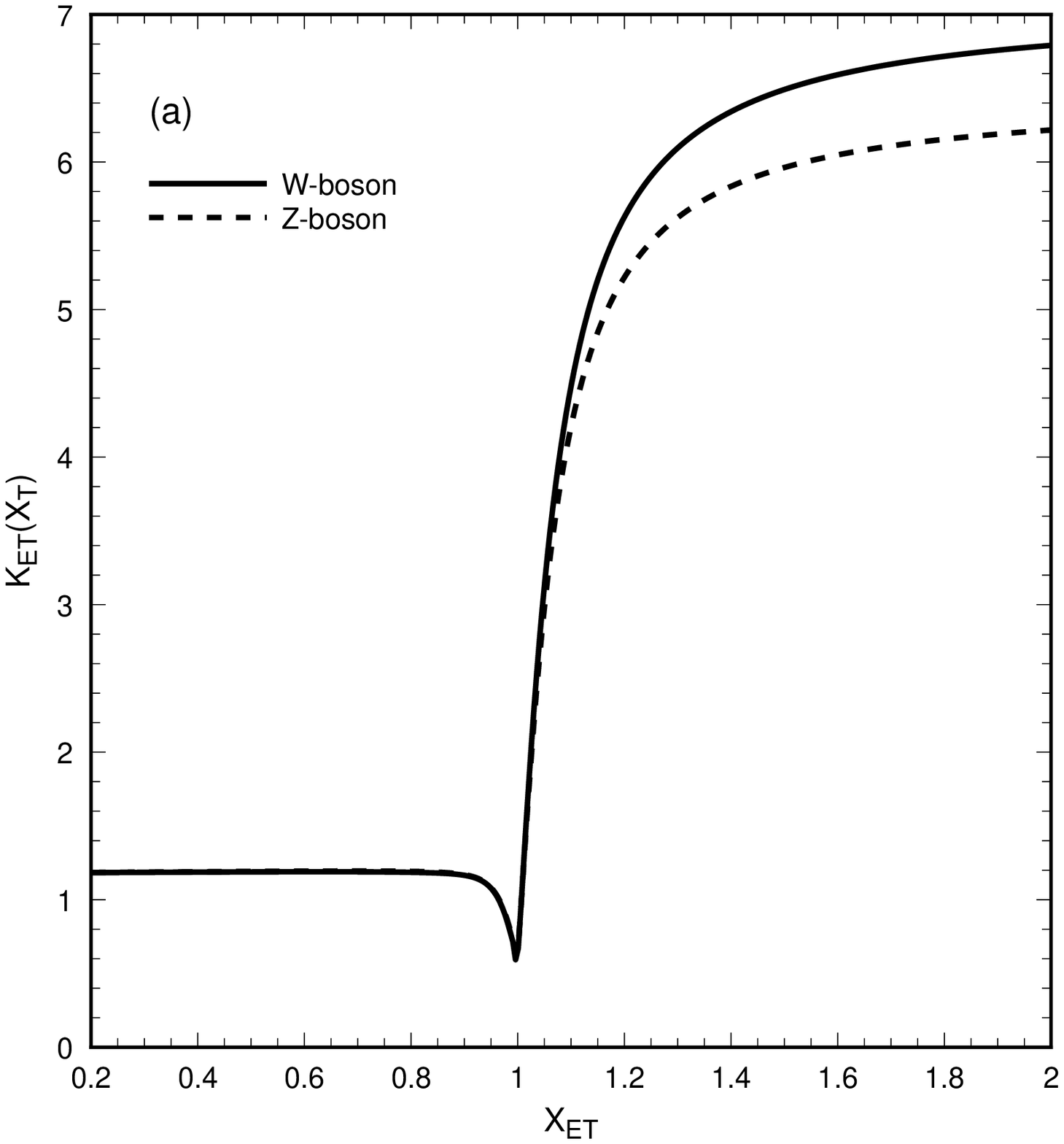}
\includegraphics{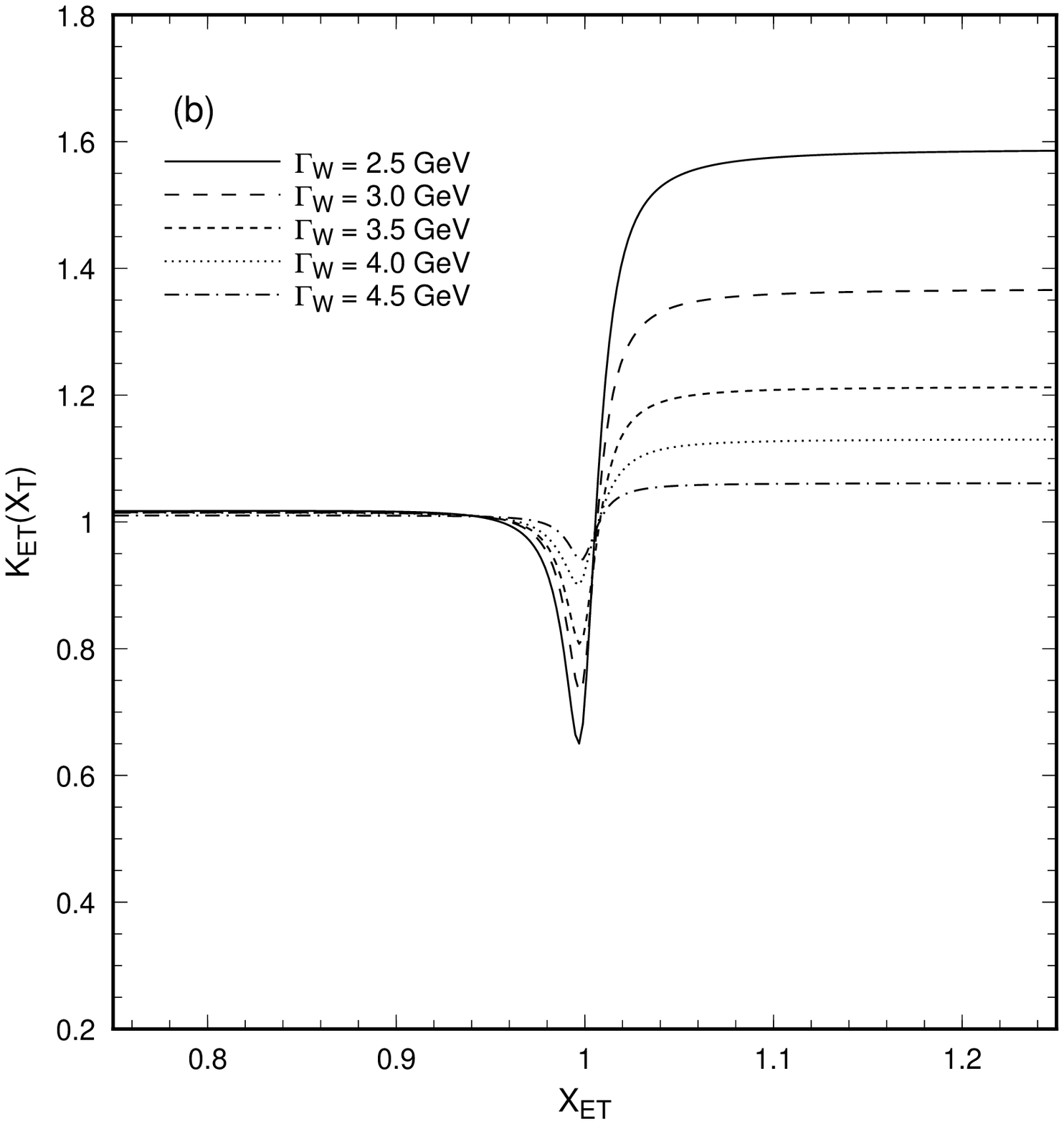}
\caption[]
{a) The $K$-factor of the lepton transverse energy distribution  
 for both the $W$-boson (solid line) and $Z$-boson (dashed line).
 b) The $K$-factor (NLO/LO) for the ratio
$\left(d\,\sigma^W/d\,X_{E_T}(\Gamma_W)\right)
/\left(d\,\sigma^W/d\,X_{E_T}(\Gamma_W = 5\ \mbox{GeV})\right)$ for different
$\Gamma_W$ in the numerator.}
\end{figure} 
\begin{figure}[t]\vspace{10cm}
\includegraphics{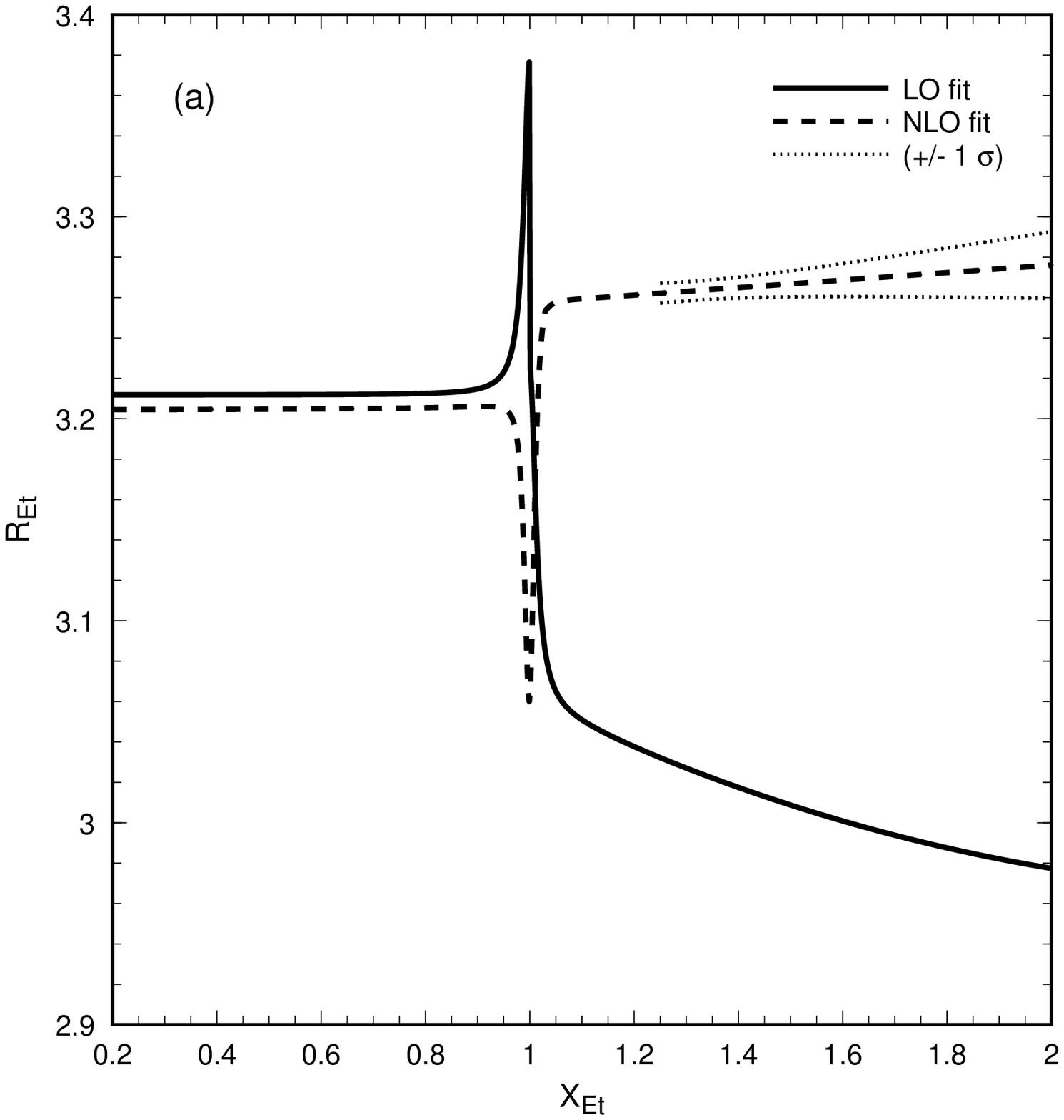}
\includegraphics{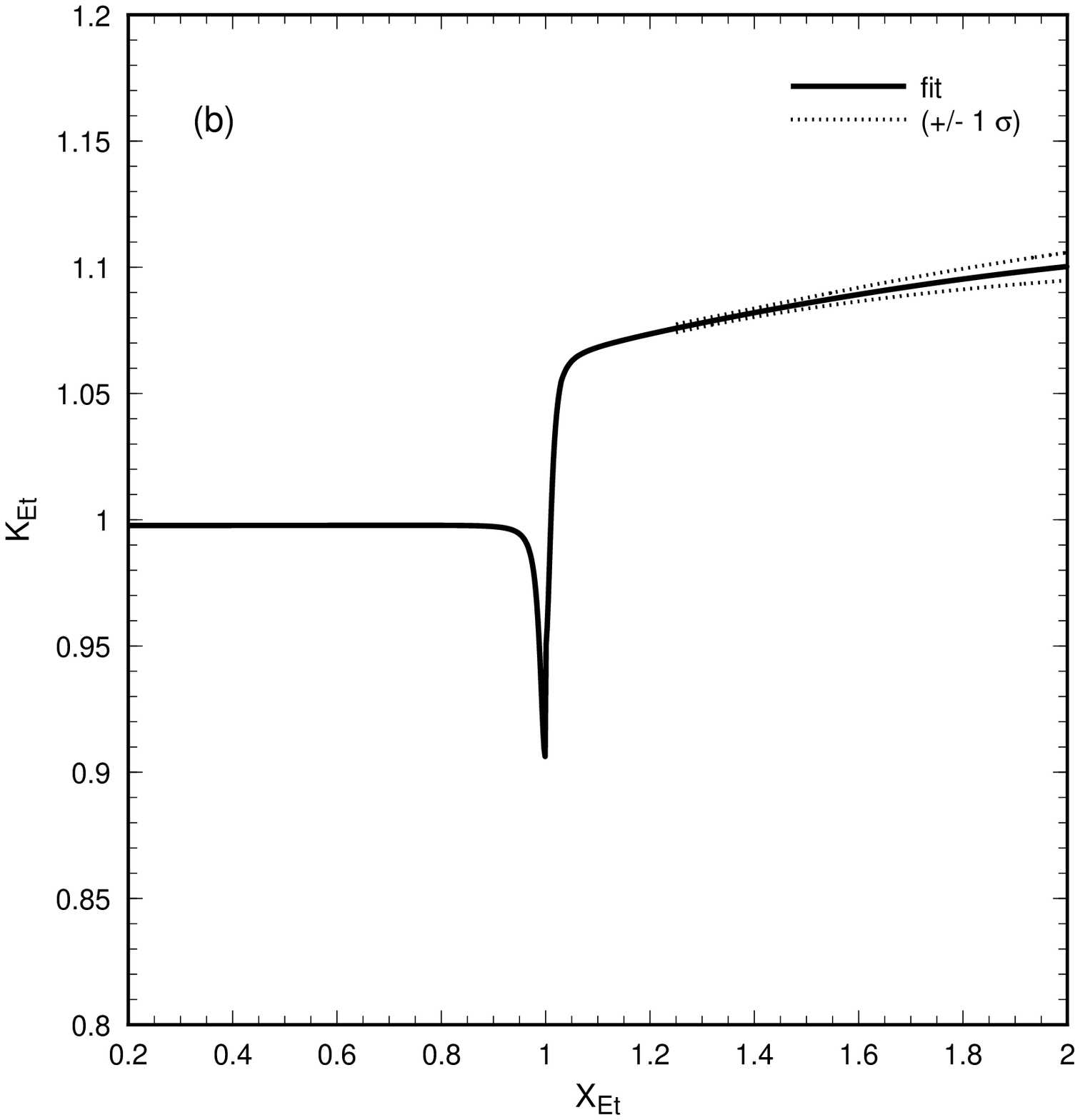}
\caption[]
{a) The LO (solid line) and NLO (dashed line) ratio $R_{E_T}$ 
as a function of the lepton scaled transverse energy $X_{E_T}=2\times E_T^V/M_V$.
b) The $K$-factor $K_{E_T}=R_{E_T}^{(1)}/R_{E_T}^{(0)}$ 
as a function of $X_{E_T}$.
The dotted lines represent the one sigma uncertainty associated
with the Monte Carlo integration.}
\end{figure}

In fig.~1a we show the vector boson $P_T$-ratio, 
$R_{P_T}(X_T)$, at three orders in $\alpha_S$~\footnote{
For the order $\alpha_S^2$ the two loops virtual corrections are not included
in the calculation, but it is enough to realize that these contributions 
factorize.}.  The lowest order ratio
$R^{(0)}_{P_T}$ is quite trivial, because $P_T^V$ is always zero due to
the lack of any initial state radiation. 
The value of the ratio is simply given by the ratio of born cross sections
\begin{equation}
R^{(0)}_{P_T} = \frac{\sigma_W^{(0)}}{\sigma_Z^{(0)}}
                   =\left( 3.216 \pm 0.001\right)\ .
\end{equation} 
The uncertainty is due to the Monte-Carlo (M.C.) integration.
The NLO ratio $R^{(1)}_{P_T}(X_T)$ at order $\alpha_S$ can have non-zero
$P_T^V$ due to initial state radiation. 
However, for the virtual
corrections the $P_T^V$-value remains identical to zero. 
This causes the low $P_T^V$-region to become infrared unstable 
and a large logarithm is generated, see eq. (7).
As given in eq.~\ref{eq:limit1}, for the ratio 
$R_{P_T}(X_T)$ this means that the 
$X_{P_T}\rightarrow 0$ limit is determined by the leading order
ratio $R^{(0)}_{P_T}(X_T)$:
\begin{equation}
\lim_{X_{P_T}\rightarrow 0} R^{(n)}_{P_T}(X_T) = R_{P_T}^{(0)}\ .
\label{eq:limit2}
\end{equation} 
This can be seen numerically in fig.~1a.
Away from $X_{P_T}=0$ there is a slow rise in the ratio
as a function of $X_{P_T}$. This is simply a consequence of the combination of
phase space effects and parton distribution functions.
The order $\alpha_S^2$ is important so we can estimate the theoretical
uncertainty in the ratio prediction. Again, close to the infrared unstable point
the behavior of $R^{(2)}_{P_T}(X_T)$ is given by eq. \ref{eq:limit2}. 
As can be seen in fig.~1b the radiative effects 
to the ratio are very small and, within the numerical
accuracy of the M.C., consistent with zero, i.e.
$|1 - K_{P_T}(X_T)| < 0.05$~\footnote{This number can be improved by simply
running the Monte-Carlo integration with a larger number of events.} 
in the relevant $X_T$-region.  Using this information we can make
our prediction for the ratio. 
The central value is $R^{(2)}_{P_T}(X_T)$ and as an 
conservative estimate of the uncertainty we take 
$|R^{(2)}_{P_T}(X_T)-R^{(1)}_{P_T}(X_T)|/R^{(2)}_{P_T}(X_T) = 
|1 - K_{P_T}(X_T)| < 5\%$ 
over the entire $X_{P_T}$ region of interest. Note that for small 
values of $X_{P_T}$ the uncertainty is 
substantial smaller ($< 2\%$) and the uncertainty 
is certainly correlated to some degree
between different $X_{P_T}$-values.
At the moment the corrections are very small 
compared to the experimental uncertainties
and can, for all practical purposes, be neglected.

In fig.~2a we show the published CDF and D0 $Z$-boson $P_T^Z$-data 
together with a fit to the D0 data (the solid line) 
and its uncertainties (the dotted lines).  All the data we are using 
were corrected for the effect of leptonic cuts.  Using
the calculated ratio shown in fig.~1a we can predict the CDF $W$-boson
$P_T$-spectrum using the measured D0 $Z$-boson $P_T$-spectrum, see 
eq.~\ref{eq:WvsZ}.  The results are given 
in fig.~2b with the order $\alpha_S^2$ ratio-prediction. 
The prediction agree very well with the $W$-boson $P_T$-spectrum for
all measured points (as low as 1 GeV). However, as expected the
order $\alpha_S$ and $\alpha_S^2$ 
prediction for the $W$-boson distribution itself 
fails below 20 GeV due to the presence of large logarithms.  
Note that the experimental
uncertainties in this particular data set is large and we do not have
to worry about the theoretical uncertainty on the ratio.  
It would be very interesting to repeat this analysis for the more recent 
data sets of
both CDF and D0 with each have an integrated luminosity over 100
pb$^{-1}$.  This would reduce the experimental uncertainties
substantially and possibly probe regions 
of smaller transverse momentum \cite{newZ}.

In some respect the method that we are suggesting here is equivalent to 
the resummed calculation.  Although the latter qualitatively describes the
$P_T^W$-spectrum, before it can make a quantitative prediction
it needs data (e.g. the $P_T^Z$-spectrum)  
to determine phenomenological non-perturbative parameters.  Therefore, 
both the ratio method and the resummed calculation can 
use the $Z$-boson data to predict
the $W$-boson data.  However our method only uses perturbative QCD, 
we do not need to resum large logarithms in the low $P_T^V$-region.  
In fact, it is in that region that the uncertainty is the smallest.
As the transverse momentum of the vector 
boson becomes smaller the QCD radiation 
becomes more and more independent of the hard process, i.e. it 
factorizes.  In the ratio the radiative corrections 
start to cancel more and more
and the ratio converges to the leading 
order ratio leaving no theoretical 
uncertainty.  If deviations are observed from 
the predicted behavior by this method
in the low $P_T$-region they can only be 
ascribed to either PDF uncertainties or to 
higher twist (non perturbative) effects. 
Because in the ratio method only perturbative QCD is used,
the estimate of the uncertainty in the theoretical prediction
is well understood.

\section{The transverse mass of the lepton pair}

The transverse mass, $M_T$,  distribution is currently used to determine the 
$W$-boson mass \cite{CDFW} and width \cite{CDFwidth1}. 
As already mentioned, it is expected that for this method 
at high luminosity the systematic uncertainty will
be larger than the statistical uncertainty~\cite{TeV33}.  
This is mainly due to the increase in the 
number of interactions per crossing and the corresponding degradation
of the neutrino transverse momentum reconstruction.  
The method described in this paper will therefore be very useful at 
high luminosity, as it trade systematic for statistical uncertainties.

In fig.~3a we show both the LO and NLO ratio $R_{M_T}$ as a function of
the scaled transverse mass $X_{M_T}=M_T/M_V$.  As can be seen
there is a remnant of the Breit-Wigner resonance at $X_{M_T}=1$.
This is due to the fact that the scaled $W$-boson width
(i.e. $\Gamma_W/M_W$) is about 10\% smaller that the scaled
$Z$-boson width, making the $W$-boson scaled distribution 
slightly narrower than the $Z$-boson scaled distribution.
Below $X_{M_T}=1$ the LO ratio tends rapidly to the LO cross
section ratio of 3.216 and is for all practical purpose a constant.
Above the resonance region the LO ratio 
is slowly falling, again due to the larger scaled $Z$-boson
width.  The radiative corrections are small, this can also be seen in fig.~3b
where $K_{M_T}$, the $K$-factor of the ratio $R_{M_T}$, is plotted.  
The range corresponds to the Monte-Carlo integration uncertainty.
In the region most
relevant for the $W$-boson mass determination ($0.9 < X_{M_T} < 1.1$)
the corrections are smaller than  0.3\% and can almost be neglected 
altogether. 
This means that the theoretical uncertainty in this region is
conservatively less than 0.3\%.  
In the tail region above 
$X_{M_T} > 1.1$, the corrections become slightly larger and positive.
The radiative corrections affect the narrower $W$-boson
transverse mass distribution more.  This region is of interest because it
can be used to determine the $W$-boson width~\cite{CDFwidth1}.  
At high luminosity~\cite{TeV33} the method 
described in ref.~\cite{CDFwidth1}
might result in a smaller uncertainty on the width than
the more traditional method using the ratio
of the inclusive $Z$- and $W$-boson cross sections \cite{CDFwidth2}.  
The method described in this paper 
should give the best
constraint on the width as it combines both the shape of the transverse mass
distribution and the total cross section ratio
between $W$- and $Z$-boson production.  
Even for large $X_{M_T}$ (up to $X_{M_T}=2$) 
the radiative corrections to the ratio are
less than 1\%. This gives us a NLO prediction for the radiative tail region
with a conservative theoretical uncertainty of less than 1\%.
Clearly there is no need for resummation in the perturbative calculation of
this observable.  

The fact that the corrections to the ratio are very small is not surprising.
It is well known that the transverse mass distribution itself
has small radiative corrections, i.e. is not sensitive to the 
$P_T^W$-spectrum.  This was the very reason,
the authors of ref. \cite{NeVe} suggested this
particular distribution as a sensible observable 
to measure the $W$-boson mass.  
In fig.~3b, we have also plotted the $K$-factor 
of the $W$-boson transverse mass
distribution itself, normalized to 1 at $X_{MT}=1$.  
As can be seen, the corrections to 
the shape of the distribution are of the order of a few percents, 
even though the overall size of the corrections are of the order of 20\%.
The advantage of using the ratio method is noticeable.

\section{The leptonic transverse energy distribution}

The transverse energy distribution of the charged lepton 
is subject to large radiative corrections.  
To use this distribution to extract the $W$-boson
mass must involve a rigorous understanding of the
large corrections and the correlation of these
corrections with the $W$-boson mass and width.
The large NLO corrections are illustrated in
fig.~4a where the $K$-factor is presented for both the 
$W$- and $Z$-boson cases as a function of
$X_{E_T}= 2 E_T/M_V$~\footnote{We normalize this scaled variable such that
the remnant of the jacobian peak is at $X_{E_T}=1$.}.  The
radiative corrections are about 40\% at $X_{E_T} = 1$ and rise
to over 600\% for $X_{E_T} > 1.1$.   Clearly, the
perturbative prediction of the shape of this distribution is unreliable.
This is unfortunate, because for this distribution it is obviously 
not necessary to reconstruct the transverse energy of the neutrino.
However, the ratio method significantly reduces the theoretical 
uncertainty and will make this distribution useful for extraction of 
the $W$-boson mass and width.

The LO and NLO ratio $R_{E_T}$ are presented in fig.~5a as a 
function of $X_{E_T}$.
At LO the ratio is identical to the LO transverse mass ratio (fig.~3a).
Below the jacobian peak region ($X_{E_T} = 1$), the NLO ratio
is very close to the LO ratio.  
However, around and above the resonance region the 
shape of the NLO ratio is inverted compared to the LO ratio.
The radiative corrections increase the width 
of the $W$- and $Z$-boson distribution
by about the same amount.  This result into a scaled width of the distribution
which is bigger in the $W$-boson case 
than for the $Z$-boson case, as $M_W < M_Z$,
resulting in the inverted shape.  
In the transverse mass case (and for that matter the invariant mass case)
the radiative corrections do not change the width of the distribution 
because of the correlation between transverse momenta 
of the two vector boson decay leptons.

Even though the corrections are larger than in the transverse 
mass case, by taking the ratio we have achieved a cancelation of the 
radiative corrections by about an order of magnitude.    
Around (above) the resonance region, there is a shape (normalization) 
uncertainty of at most 10\% as can been seen in fig. 5b.
As the experiments become more accurate it might be necessary
to extend the calculation of the ratio one order higher in
$\alpha_S$ and thereby reducing the theoretical uncertainty 
substantially.  All the virtual contributions needed are already
known in the literature~\cite{Neerven} and such a calculation 
is certainly within the realm of possibilities. 
Using the ratio method for the lepton transverse energy 
with the currently available  CDF and D0 data containing over
100 pb$^{-1}$ will tell us if such a
calculation is necessary.

It is interesting to note that the cancellation of the large corrections 
is due in part to the closeness of the scaled widths.
To demonstrate the relation between the width and the size
of the corrections we show in
fig.~4b the NLO $K$-factor for the ratio
$\left(d\,\sigma^W/d\,X_{E_T}(\Gamma_W)\right)
/\left(d\,\sigma^W/d\,X_{E_T}(\Gamma_W = 5\ \mbox{GeV})\right)$ for different
$\Gamma_W$ in the numerator.  Clearly if the scaled widths were not close, 
the corrections to the ratio would still be large.

\section{Conclusions}

In this paper we have presented an alternative method to calculate
$W$-boson observables by using the experimentally measured $Z$-boson
observable and a perturbative calculation of the ratio of the $W$- over
$Z$-boson observables.  The radiative corrections, which can be large
for these observables, tend to cancel in the ratio.
Using this method, the predicted
transverse momentum distribution and the transverse mass distribution 
of the $W$-boson have small radiative corrections
over the whole range of relevant values.
For the transverse energy distribution of the lepton a NNLO prediction
might be necessary for the main injector data. 
With the current data
sets the method described in this paper can be used to 
estimate the future expected uncertainties. One can then
decide on the necessity of higher order predictions.
It is worthwhile to investigate 
the method outlined in this paper
as an alternative
way to measure the $W$-boson parameters.
Possibly, this method could augment/replace the current
methods at high luminosity runs. It reduces 
both theoretical and experimental systematic
uncertainties at the price of increasing the statistical 
uncertainty. Such a trade-off becomes
more and more interesting as the
integrated luminosity increases at the TEVATRON.


\begin{thebibliography}{99}
\bibitem{BeKl} F.~A.~Berends, R.~Kleiss, J~.P.~Revol and J.~P.~Vialle,
               Z.~Phys.~C27 (1985) 155. 
\bibitem{BaKeWa} ``Electroweak radiative corrections
                 to $W$- and $Z$-boson production in hadronic collisions'', \\
                 U.~Baur, S.~Keller, W.~Sakumoto and D.~Wackeroth,
                 UB-HET-96-02, hep-ph/9609315,
                 to appear in the proceedings of the 1996 
                 Annual Meeting of the Division 
                 of Particles and Fields (DPF 96) 
                 of the American Physical Society, 
                 Minneapolis, MN, 10-15 Aug 1996.  \\
	 	 ``Electroweak Radiative Corrections to $W$-Boson Production
                 at the Tevatron'', U.~Baur, S.~Keller and D.~Wackeroth,
		 to appear in the Proceedings of the 1996 DPF/DPB Summer 
                 Study on New Directions for High-Energy Physics 
                 (Snowmass 96), Fermilab-Conf-96/424-T, UB-HET-96-04.  
\bibitem{D0W} The D0 Collaboration, Phys.~Rev.~Lett.~77 (1996) 3309.
\bibitem{LaYu} G.~A.~Ladinsky and C.~P.~Yuan, Phys.~Rev.~D50 (1994) 4239.
\bibitem{AR} P.~B.~Arnold and M.~H.~Reno, Nucl.~Phys.~B319 (1989) 37;
             ERRATUM-ibid.~B330 (1990) 284;\\
             P.~B.~Arnold, R.~K.~Ellis and M.~H.~Reno, 
             Phys.~Rev.~D40 (1989) 912.
\bibitem{AK} G.~Altarelli, R.~K.~Ellis and G.~Martinelli, 
             Nucl.~Phys.~B143 (1978) 521;\\
             C.~T.~H.~Davies, B.~R.~Webber and W.~J.~Stirling,
             Nucl.~Phys.~B256 (1985) 413;\\
             P.~B.~Arnold and R.~P.~Kauffman, Nucl.~Phys.~B349 (1991) 381;\\
             ``Vector boson production in hadronic collisions'',
	     R.~K.~Ellis, D.~A.~Ross and S.~Veseli,
             Fermilab-pub-97/082-T.
\bibitem{CoSoSt} C.~Collins, D.~E.~Soper and G.~Sterman, Nucl.~Phys.~B250 
                 (1985) 199. 
\bibitem{CDFW} The CDF Collaboration, Phys.~Rev.~Lett.~75 (1995) 11.
\bibitem{CDFWFULL} The CDF Collaboration, Phys.~Rev.~D52 (1995) 4784.
\bibitem{TeV33} ``Future electroweak physics at the Fermilab TEVATRON: 
                report of the TEV-2000 study group'', D.~Amidei {\it et al.}, 
                FERMILAB-PUB-96-082.
\bibitem{lepZ} The ALEPH Collaboration, the DELPHI Collaboration, 
               the L3 Collaboration and the OPAL Collaboration, 
               CERN-PPE-94-187, contributed to the 27th International
               Conference on High-Energy Physics - ICHEP 94, 
               Glasgow, Scotland, UK, 20 - 27 Jul 1994. 
\bibitem{Reno} M.~H.~Reno, Phys.~Rev.~D49 (1994) 4326.
\bibitem{Neerven} T.~Matsuura and W.~L.~van Neerven, Z.~Phys.~C38 (1988) 623;\\
                  T.~Matsuura, S.~C.~van der Marck and W.~L.~van Neerven,
                  Phys.~Lett.~211B (1988) 171; Nucl.~Phys.~B319 (1989) 570;\\
                  T.~Matsuura, R.~Hamberg and W.~L.~van Neerven, 
                  Nucl.~Phys.~B345 (1990) 331; Nucl.~Phys.~B359 (1991) 343.
\bibitem{SnoPDF} There were some discussions around the issue of 
PDF uncertainties during Snowmass 96. 
\bibitem{CDFWPt} The CDF Collaboration, Phys.~Rev.~Lett.~66 (1991) 2951.
\bibitem{D0ZPt} ``Measurement of the $Z$-boson transverse
    momentum distribution in $p\bar p$ collisions at 
    $\sqrt{S}=1.8$ TeV with the D0 detector'', Zhi-Yu Jiang, 
    Aug 1995, Ph.~D.~Thesis, 
    State University of New York at Stony Brook.
\bibitem{CDFwidth1} The CDF Collaboration, 
Phys.~Rev.~Lett.~74 (1995) 341.
\bibitem{CDFwidth2} The CDF Collaboration, 
Phys.~Rev.~Lett.~73 (1994 ) 220; Phys.~Rev.~D52 (1995) 2624;\\
The D0 Collaboration, Phys.~Rev.~Lett.~75 (1995) 1456.
\bibitem{dyrad} W.~T.~Giele, E.~W.~N.~Glover and D.~A.~Kosower,
                Nucl.~Phys.~B403 (1993) 633.
\bibitem{CDFZPt} The CDF Collaboration, Phys.~Rev.~Lett.~67 (1991) 2937.
\bibitem{newZ} ``$P_T$-dependence of inclusive $Z$-boson production'',
D.~P.~Casey, for the D0 Collaboration, to appear in
the proceedings of the DPF96 Conference, Minneapolis, 
MN, August~10 --~15, 1996,
Fermilab CONF-96/272-E. 
\bibitem{NeVe} J.~Smith, W.~L.~van Neerven and J.~A.~M.~Vermaseren,
               Phys.~Rev.~Lett.~50 (1983) 1738.
\bibitem{RATIO} 
``$M_W$ measurement at the TEVATRON with high luminosity'',
W.~T.~Giele and S.~Keller, to appear in the 
proceedings of the DPF96 Conference, Minneapolis, MN, August~10 --~15, 1996,
FERMILAB-CONF-96/307-T;  \\
``Measurement of $M_W$ using the transverse mass ratio 
of W and Z'', 
S.~Rajagopalan and M.~Rijssenbeek, for the D0 Collaboration, 
to appear in the Proceedings of the 1996 DPF/DPB Summer Study on New 
Directions for High-Energy Physics (Snowmass 96),  FERMILAB-CONF-96-452-E.
\end{thebibliography}
\end{document}